\documentclass[12pt, preprint]{aastex}

\newcommand{\eq}{\begin{equation}}
\newcommand{\ef}{\end{equation}}
\newcommand{\eqa}{\begin{eqnarray}}
\newcommand{\efa}{\end{eqnarray}}

\begin{document}
\title{Cloaked Gamma Ray Bursts}
\author{David Eichler\\
        Physics Department, Ben-Gurion University, Be'er-Sheva 84105, Israel\\
        E-mail: \email{eichler.david@gmail.com}}

\begin{abstract}
It is suggested that many $\gamma$-ray bursts (GRBs) are cloaked by an ultra-relativistic baryonic shell that has high optical depth when the photons are manufactured. Such a shell would  not fully block photons reflected or emitted from its inner surface, because the radial velocity of the photons can be less than that of the shell. This avoids the standard problem associated with GRBs that the thermal component should be produced where the flow is still obscured by high optical depth. The radiation that escapes high optical depth obeys the Amati relation.
Observational implications may include a) anomalously high ratios of afterglow to prompt emission, such as may have been the case in the recently discovered PTF 11agg, and b) ultrahigh-energy neutrino pulses that are non-coincident with detectable GRB. It is suggested that GRB 090510, a short, very hard GRB with very little afterglow, was an {\it exposed} GRB, in contrast to those cloaked by baryonic shells.
\end{abstract}

\section{Introduction}
A longstanding puzzle concerning GRBs 
is that their highly super-Eddington luminosities would seemingly imply that the very photons that comprise the GRB should drag out enough baryonic material to obscure themselves.
The problem is compounded by the discovery that long GRB occur inside massive stars, meaning that the GRB fireball must plow up material from  its host star as it pushes its way out, not all of which can move out of its way. Neutrons drifting into the fireball ( Eichler \& Levinson, 1999, Levinson \& Eichler, 2003), and highly opaque pre-collapse winds from the host stars (Ofek et al., 2013) may also cloak GRBs with an optically thick layer out to $10^{3.5}$ lightseconds.  Naively, this would seem to be enough to completely hide the GRB, which typically lasts less than $10^2$ s.

One suggestion to solve the first problem is that the energy powering the GRB originates on magnetic field lines that thread a black hole that forms at the center of the  collapsing host star (e.g. Levinson and Eichler, 1993), which should be nearly devoid of baryonic material, but this does not address the issue of the plowed-up host star material, which cloaks the fireball even on  horizon-threading field lines.

  One solution to the cloaking problem is that many GRBs are seen by observers just off the (angular) edge of the jetted fireball  (Eichler \& Levinson, 2004, Levinson \& Eichler 2006), and that this accounts for the Amati and Ghirlanda relations  (Amati, 2002, Ghirlanda  et al 2004 ). The hypothesis successfully predicted an intermediate flat phase for GRB afterglows (Eichler, 2005, Eichler \& Jontof-Hutter, 2005) that was later confirmed by Swift observations.
   The "off-edge observer" hypothesis also solves the problem\footnote{See Vurm, Lyubarsky \& Piran (2013) for a detailed, carefully quantified discussion of this point. They do not, however, consider scattering within the host star envelope off collimating walls, which could lower the photon energies in the observer frame via Compton recoil and pair production.} that because blue-shifted $\gamma$-ray photospheres should have spectral peaks that are or order $3 \Gamma kT_{ann}\gtrsim 1$ MeV, where $kT_{ann}$, the temperature at which pairs annihilate and permit optical transparency, is of order 10 KeV, and $\Gamma$, the bulk Lorentz factor of the flow, is $\gtrsim 10^2$,  the spectral peak should then be $\sim 3$ MeV, whereas most are observed to have spectral peaks at $\lesssim 300$ KeV.  Taking into account the possibility that pair production opacity limits the photon energy to $\sim 1$ MeV, there is still some discrepancy with the fact that the distribution of $E_{peak}$ for {\it long}  GRB (in contrast to short, hard GRB) peaks  comfortably below the pair production threshold. Because the  hypothesis  predicts that the observed spectral peak is softened relative to that seen by a  head-on observer ($\theta=0$) by a factor of $S(\theta)\equiv E_{peak}(0)/E_{peak}(\theta) ={(1-\beta \cos \theta)\over(1-\beta)}=(1 + {\beta(1-\cos \theta)\over(1-\beta)})\sim  1+(\theta\Gamma)^2$, where $\theta$ is the viewing offset angle and $\Gamma$ is the bulk Lorentz factor, the problem disappears if $ \theta \Gamma \sim 3$.
    The "off-edge" viewing hypothesis, however,  does not yet  explain why {\it most} GRBs are viewed from such an offset angle, especially considering that the opening  angle for GRB, $\sim 0.1$, is probably larger than $1/\Gamma$. (Indeed, about half of all classical GRB, although they show spectral peaks of $\lesssim 300$ KeV, do  {\it not } display a  noticeable flat phase in their afterglow.) Neither would it solve the cloaking problem if the cloak has a larger solid angle than the jet, as in the case of a thick stellar wind.

    \section{Cloaked GRB}
    In this letter, I propose an alternative to the off-edge viewing hypothesis: that the cloak becomes optically thin before the last of the GRB photons that scatter off its back end reencounter it. Thus, photons that scatter off (or are emitted by) the back end of an optically thick cloak  reach the observer as long as their direction in the observer frame is sufficiently offset from the local velocity vector of the cloaking material that the reencounter occurs only after the cloak has become transparent. This alternative is not advanced as being the case for {\it most} GRB, as at least half of long GRB have flat afterglow phases;  it is proposed specifically for those that have prompt afterglow or very short flat phases.


\begin{figure}
\begin{center}
\includegraphics[scale=0.5]{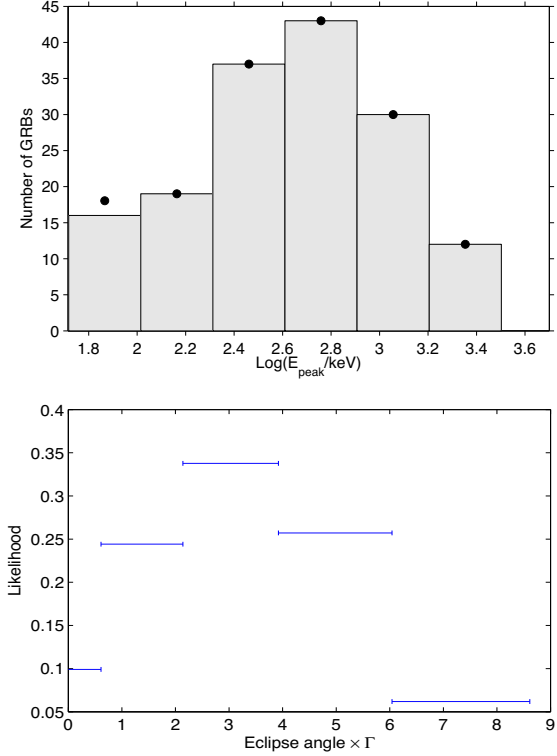}
\label{eclipsed_GRB}
\caption{The  observed number of GRBs is displayed (top panel) as a function of peak spectral energy $E_{peak}$ assuming the value at the source is 3 MeV. The predicted number of GRBs in these logarithmic bins are displayed by the dots in the top panel, for a distribution of eclipse angles $\theta_1$ that peaks at 3/$\Gamma$, (bottom panel). The eclipse angle distribution is chosen such that the observed number of GRBs in all energy bins with $E_{peak} \ge 100$ KeV (i.e all but the leftmost energy bin the top panel) is reproduced.
A value of 0.1 is assumed for the jet opening angle $\theta_o$, while the number of GRB with  $\theta_1 \ge 8.6/\Gamma$ is assumed to vanish.
 It is assumed 
 that any GRB with $E_{iso}$ above $10^{50}$ erg (i.e. $E_{peak}$ above 30 KeV according to the Amati relation) can be seen at any redshift, while those with $E_{iso}$ below $10^{50}$ erg have a negligible $V_{max}$ and are not observed. The  GRBs with observed $E_{peak}<100$ KeV, shown in the leftmost bin, is due to large viewing offset angles, i.e. observers well outside $\theta_o$, and is not a free parameter. }
\end{center}
\end{figure}

\begin{figure}
\begin{center}
\includegraphics[scale=0.5]{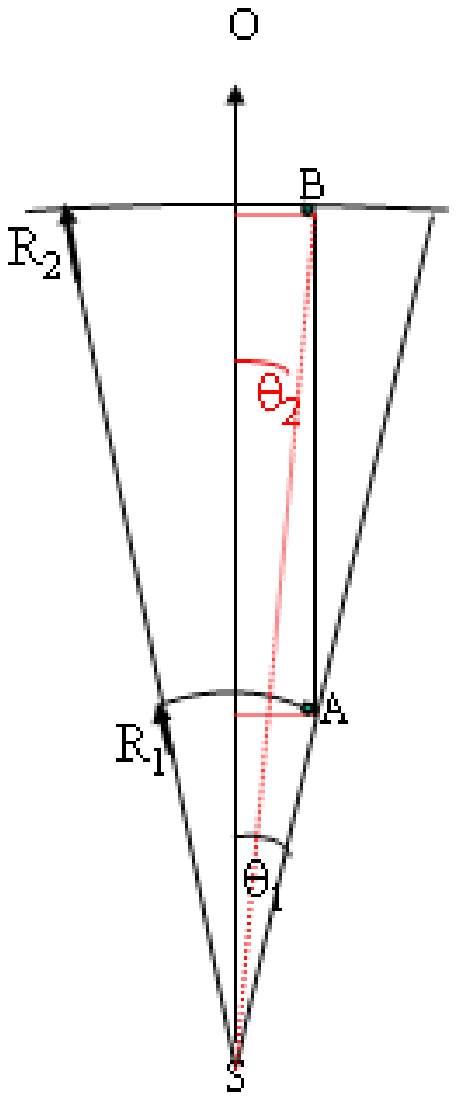}
\label{variation_d}
\caption{The photon originates at the central source S, scatters off point A on the shell, when it is at radius $R_1$,  into the direction of the observer O, and reencounters the shell at point B, when it is at radius $R_2$. If $R_2$ is  the radius at which the shell becomes transparent - then $\theta_1$ is the "eclipse " angle  within which photons scattered at $R_1$ would not reach the observer.}
\end{center}
\end{figure}


 We now consider the possibility that, when the fireball breaks through the stellar surface,  some matter is plowed up and trapped at the front of the fireball with a column mass density $\Sigma m_p/\sigma_T \equiv 10^7\Sigma_7 m_p/\sigma_T $.  
 A GRB with an 
 $E_{iso}$ of $10^{54}E_{iso,54}$ erg  can  eventually push this column to a Lorentz factor of order
 \begin{equation}
 \Gamma_a
\sim  E_{iso}\sigma_T/4\pi R_{}^2 \Sigma m_p c^2 
 \sim 10^{2.5} E_{iso,54}/ R_{11}^2 \Sigma_7
 \label{Gamma}
 \end{equation}
 {
   Hence $R_{11}^2\Sigma_7 \lesssim 1$ is allowed by the empirical condition that $\Gamma _a\gtrsim 10^{2.5}$, which may  be a preferentially selected range for  {\it observable} GRB. We do not claim to calculate from first principles the value of $\Sigma$. We may  estimate $\Sigma$ by assuming it to be 
   the surface density at which the acceleration time $t_{acc} =(L/L_{Edd})^{-1}(Rc^2/GM)\Sigma \Gamma^{3}R/c$ would be of the order of $t_{acoustic}\equiv \sqrt3 R\theta_o/c$, as slower acceleration would allow $\Sigma$ to rapidly decrease, which hastens the acceleration.\footnote{Thompson (2006) equates the acoustic time with the proper hydrodynamic time $R/\Gamma c$.} For $\theta_o\sim 0.1$, $E_{iso, 54}\sim 1$, this suggests   $\Sigma \sim 10^7$.

A thin baryonic shell of grammage $\Sigma(R)m_p/\sigma_T$ that is trapped ahead of the fireball at the moment of breakout from the host star remains optically thick out to a radius $R_2$ that is determined by the condition $R_2 =  R_1[\eta(R_2)\Sigma(R_1)]^{1/2}$, where $\eta$ is the fraction of its thickness $\Delta$ traversed by the photon over an expansion time. If the thickness is less than $R(1-\beta)$, the light, which moves at a relative velocity (as seen by the observer) of $(1-\beta) c$  to the shell, has enough time to completely cross the shell over one expansion time, so it must traverse the full optical depth of the shell, and $\eta =1$. Otherwise $\eta \sim R(1-\beta)/\Delta$.   

Because the shell is  ultrarelativistic, photons scattering {\it backward} off its inner side can nevertheless reach the observer.
Assume the shell expands spherically. A diagram is given in figure  2  where point A is at coordinates $ (R_1,\theta_1)$, where $\theta$ is the  angle from  the observer's line of sight, and B is at $(R_2, \theta_2)$. A photon from the central engine that is scattered towards the observer from  point A  on the inner side of the shell reencounters the shell at point B  - after the shell has expanded to  radius $R_2$ - if  the elapsed time $\delta t_{1,2}= D_{AB}/c$ required for the photon to travel the distance $D_{AB}$ from A to B equals the time $\delta t_{1,2} = \int_1^2 dR/\beta(R) c  \equiv (R_2 -R_1)\overline{\beta^{-1}}/c$ required by the shell to expand from $R_1$ to $R_2$. If this condition is first met after the shell has become optically thin, then the photon probably escapes the shell and reaches the observer. Otherwise it  either rescatters off the inner side of the shell or, if the shell is only a polar cap, escapes around the shell's edge. (Note that it can do both - i.e. repeatedly scatter and eventually make its way to beyond the edge.)  As the photon is scattered  from radius $R_1$ at a   point that makes an  angle $\theta_1$ from the viewing axis, it must scatter by  $\theta_1$  from the radial direction into the observer's line of sight, and the distance AB is given by
 \eq
 D_{AB}= R_2\cos \theta_2 -R_1\cos \theta_1,
 \ef
 where $\theta_2$ is given by $R_2\sin \theta_2 = R_1 \sin \theta_1$.
 So the condition for the photon to catch up to the shell at radius $R_2$ is then \eq
 (1-\cos \theta_1) = (R_2/R_1 - 1)(\overline{\beta^{-1}}-1)+ (R_2/R_1)(1-\cos \theta_2).
\ef
When $\Sigma(R_1) \gg 1$, $R_2 \gg R_1$,  the second term on the right hand side is negligible compared to the first term. If in addition ${\beta^{-1}}\equiv 1+
1/2{\Gamma^2}$ is constant, then $\theta_1^2 \simeq (R_2/R_1) \Gamma^2 $.

Now suppose that, at radius $R_1$, the baryonic shell has an optical depth $\Sigma(R_1)\gg 1$, so that it becomes transparent only  after expanding by a factor $R_2/R_1 = \Sigma^{1/2}(R_1)$. In order to escape through the shell at $R_2$ after first  scattering off the shell at $R_1$, the shell needs to be transparent at $R_2$ and the  scattering should satisfy
\eq
 1 - \cos \theta_1 \ge (\overline{\beta^{-1}}-1)(\Sigma^{1/2}(R_1) -1 ) .
\ef
This condition defines a minimum "eclipse angle"  below which the incident photons, assuming they move radially, are obscured if they are scattered into the observer's line of sight. In figure \ref{eclipsed_GRB}, this is quantified. It is shown there that the distribution of peak energies in GRB with known redshifts can be accounted for, assuming the source $E_{peak}$ is $\sim 3 MeV$,  by  a suitable distribution of eclipse angles for randomly distributed observer lines of sight and a jet opening angle of 0.1.

Suppose $\Gamma(R) \propto R^{\alpha}$, then $ (R_2/R_1 - 1)(\overline{\beta^{-1}}-1)\sim (1-\beta(R_1))[(R_2/R_1)^{1-2\alpha} - 1] /(1-2\alpha)$. For $\alpha = 1/3$ all the way to to $R_2$ (Thompson, 2006) (though there is no guarantee that the burst lasts long enough that the radiation pressure is maintained until $R_2$, where the shell becomes transparent), it follows that
\eq
S(R_1) \equiv  {{ 1 - \beta(R_1)\cos \theta_1}\over{1-\beta(R_1)}} \ge 3[\Sigma(R_1)^{1/2}-1]^{1/3}+1
\label{ecl}
\ef

The weak dependence on $\Sigma(R_1)$ means that the observed photons that have scattered off the shell over the last few expansion times before it became transparent are softened by a modest factor of order 3 or slightly more, which is in accord with observations.

Any  photons escaping the stellar surface at $R_s$ more than
 $R_s^{2\alpha}[\beta^{-1}(R_s)-1][ R_2^{(1-2\alpha)}-R_s^{(1-2\alpha)}]/c(1-2\alpha)$ after the fireball first emerges can escape scatter-free directly toward the observer . Note, however, that they arrive before the soft, scattered photons that leave the shell at the same time.
If $R_2=\Sigma^{1/2}R_1$ is sufficiently large that equation (3) is not satisfied for any point of first scattering,
then the burst is entirely obscured even at the second scattering.


This result should be compared to the case where the photons are subjected to the opacity of a continuous outflow rather than a thin shell:  In this case, under the assumption of constant $\Gamma$, the optical depth $\tau$ is given by $\tau = ct_{ex}^{\prime} n^{\prime}\sigma_T$, where $t_{ex}^{\prime}$ is the proper expansion time $R_1/\Gamma c$ and $n^{\prime}$ is the proper density $n/\Gamma$, and the dimensionless column density is $\Sigma = n(R_1) R_1\sigma_T $. In this case, transparency obtains when $\Gamma \ge \Sigma^{1/2} $, more or less independent of the direction of the photon. {\it
So in the case of ultrarelativistic baryonic outflow, a GRB is less obscured by a thin, highly opaque shell than by a continuous outflow of the same column density.} This additional transparency applies only to photons moving backward in the frame of the shell, so the element on the shell that scattered the photons into the observer's line of sight would have had to be moving at an angle of at least $1/\Gamma$ relative to the observer's line of sight, and possibly more, depending on the actual optical depth of the shell.

Consider the not unlikely parameters for  a GRB: $E_{iso, 54}=1$, $R_{11} = 3$, $ \Gamma_a = 300$. Using equation (\ref{Gamma}), we estimate $\Sigma(R_s) = 10^6$, which is not too different from the estimate above. Even if the photons traverse only $(1-\beta)$ of the shell over its expansion time, it is still opaque to them.  On the other hand, equation (\ref{ecl}) implies that the shell allows a backward photosphere with a softening factor S  of order $30$ even for the first photons that encounter the shell.  At later stages, as $\Sigma(R)$ decreases, the eclipse angle $\theta_{min}$ and hence $S(\theta_{min})$ are reduced somewhat.

The above equations have assumed spherically symmetric shell. This is probably oversimplified, as the Rayleigh-Taylor  instability should created plumes with opening angles $\theta_p \ll \theta_o$. The head of each plume can shed matter until the Lorentz factor of the plume head, $\Gamma_p$, is of order $1/\theta_p$,
at which point the time required for further mass shedding $t_{accoustic}$  is below the proper expansion time. The individual subpulses in GRBs, which can last less than 1 second, may  be plumes such as this that point within  $\theta_v\sim \theta_p$ of the line of sight.  Although the evolving plume's structure and thickness profile are  beyond the  scope of this paper,  the condition that $\Gamma_p \theta_p \sim 1$ suggests that, as the thinnest parts of the shell first become transparent, the typical softening factor S is  $S\sim{ (1-\beta \cos \theta_v)\over (1-\beta)}$, i.e. that it is of order several, even if much of the shell is still opaque. Attributing the typically observed hard-to-soft evolution of GRBs to the acceleration of the scattering material (Eichler \& Manis, 2007, 2008, Eichler, Guetta, \& Manis, 2009) quantitatively accounts for many of the empirical scaling laws and correlations provided by observations (Ryde, 2004,  Norris \& Bonnel, 2006), and for soft X-ray tails that often accompany short hard GRB .

Photons that are cloaked by the baryons have energies far less than $m_e c^2$ in the frame of the baryonic shell, so they reemerge from the surface after several scatterings,  with little energy loss from Compton recoil,  and repeat this until the baryons finally become optically thin or until they have diffused to the side of the baryonic cap. In the latter case, their energy is concentrated into an annulus just outside  the opening  of the fireball. So the bias in favor of the viewer being  just a bit offset from the velocity vector $\hat v$ of the emitting material is the result of both a) the eclipse in the $\hat v$ direction and b) the accumulation of photons propagating just outside the cone of the fireball.   The extent of this effect depends upon the detailed parameters of the system and the history of its acceleration so we cannot offer a firm quantitative prediction.

\section{Discussion}
When the prompt $\gamma$-rays are strongly obscured, there still remain several possible observational signatures: a) a smaller, softer burst at low $E_{peak}$, b) an anomalously strong afterglow to prompt $\gamma$-ray ratio, c) an early "orphan" afterglow ( a special case of {\it b}) and d) an anomalously strong neutrino to prompt $\gamma$-ray ratio. We consider each of these.

{\it Weak prompt emission with strong afterglow}:  A sufficiently soft burst is usually termed an X-ray flash, and X-ray flashes, which respect the Amati relation, are sub-luminous relative to classical GRBs. An X-ray flash that is really a cloaked classical GRB might therefore have an anomalously high afterglow-to-prompt emission  ratio.    On the other hand, the absence of such objects would imply only that the prompt X-ray emission was sufficiently obscured that it was not observed by burst alert telescopes. The possibility still exists of searching for early orphan X-ray afterglow with wide  X-ray cameras. Such objects could be interpreted as cloaked GRBs if they decay without a flattened or depressed early phase that would be attributable to off-edge viewing.

 The unidentified  transient PTF 11agg (Cenko et al 2013) is a classic afterglow that has no association with a known GRB. The transient was discovered via its optical afterglow, which was typical of a bright GRB. Obviously, we should entertain the possibility that it was the afterglow from a cloaked GRB. While it is difficult to prove there was no GRB from one event, the authors argue that, given the known GRB rate, its discovery would have been {\it a priori} unlikely if there was an associated GRB.  In any case,  the discovery shows that "orphan" afterglows can be found without the alert of a $\gamma$-ray detection. The discovery of several more should afford enough statistics to see if {\it a posteriori} $\gamma$-ray detection is commensurate to sky coverage by $\gamma$-ray detectors or whether some of them are truly $\gamma$-ray quiet.

 Finally, we note that the very hard burst (even as compared to other short, hard GRBs)  GRB 090510, which had a spectral peak above 4 MeV, displayed very little afterglow. In the first two years of GBM operation, there were 20 GRB  of known redshift that were detected by both GBM and Swift. GRB 090510 had by far the highest $E_{peak}$, the least luminous 24 hour X-ray afterglow, and the smallest ratio of 24 hr fluence to prompt fluence. Defining the 24 hr afterglow fluence to be the 24 hr flux times 24 hours, this ratio was only  $10^{-3.5}$, whereas if the baryons that powered the afterglow also powered the prompt emission, as in the internal shock model, then this ratio should have been of order unity or greater. Why should the spectral peak and the 24 hr afterglow flux be so dramatically {\it anti}-correlated? The considerations discussed here account for this anti-correlation in simple fashion: We suggest that GRB 090510 is an {\it exposed} GRB - the exception that proves the rule that most GRB observed head-on are cloaked by a baryonic shell -  and that the hardest prompt photons from the GRB reached us without being obscured by a baryonic cloak. We also note that if GRB 090510 had been sufficiently cloaked or misdirected off our line of sight,  its $E_{peak}$ would have been 200 KeV, typical of a long GRB, and its duration would have appeared to have been over 20 seconds, due to kinematic effects,  rather than the observed duration of 1 s. The lack of cloaking may be typical of short GRB, whose progenitors lack an extended envelope.

The common element in all of the above is that the ratio of prompt $\gamma$-ray emission to other GRB-related emission can be highly variable once the possibility of partial $\gamma$-ray eclipse is considered.

 {\it Neutrinos}:    Similarly, the cloaking of the prompt $\gamma$-rays would not block any neutrinos that might be emitted,  which in fact would achieve their highest flux where the baryonic cloak would
 serve as a medium for the internal shocks that accelerate the energetic primaries (as well as the  target material for the neutrino production). Currently, no neutrinos have been detected in association with GRBs, which, as argued in Eichler (1994), is not surprising by a simple energetics argument:  the detection threshold above 1 TeV is an energy flux threshold, about $10^{-4}$ erg/cm$^2$, and the energy flux in $\gamma$-rays of even the brightest GRB is only comparable to this threshold. Moreover, Fermi LAT measurements show that the {\it non-thermal} component of the strongest bursts contributes only a small fraction of the total flux. On the other hand, the present hypothesis suggests that because most observers of GRB are off edge, we may be missing the neutrino bursts. Those GRB with the strongest neutrino bursts may be sufficiently obscured by a baryonic cloak that we miss the GRB. We therefore suggest that neutrinos detected by underwater and under-ice high energy neutrino detectors be followed up by X-ray afterglow searches within several hours.

 To summarize, this paper and references therein propose that many features of GRB can be understood as a $\gamma$-ray echo from the inside of an expanding, ultrarelativistic shell.

  I thank  S.B. Cenko, N. Gehrels, Y. Lyubarsky, C. Kouveliotou, Y. Kaneko, and E. Ofek for very helpful discussions. I thank Prof. L. Amati for sharing his data with me prior to publication, and R. Kumar for preparing the graphs. This research was supported by the Israel-US Binational
Science Foundation, the  Israel Science Foundation, and the Joan and Robert Arnow Chair of Theoretical
Astrophysics.

\centerline{REFERENCES}

\noindent Amati, L., et al. 2002, A \& A, 390, 81

\noindent Cenko, S. B., et al., 2013 arXiv:1304.4236v

\noindent Eichler. D., 1994 ApJS,  90, 877

\noindent Eichler, D. \& Levinson A., 1999 ApJ, 521, L 117

\noindent Eichler, D. \& Levinson A., 2003, 594, L19

\noindent Eichler, D., 2005, ApJ  628L 17

\noindent Eichler, D. \& Manis, H. 2007, ApJ. 699, L65

\noindent Eichler, D. \& Manis, H. 2008, ApJ. 689, L85

\noindent Eichler, D. Guestta, D. \& Manis, H. (2009) ApJ, 690, L61

\noindent Ghirlanda, G., Ghisellini, G., \& Lazzati, D. 2004, ApJ, 616, 331

\noindent Lazzati, D.   Morsony, B.J., Begelman, M.C.   2009, ApJ,  700, L47

\noindent Levinson, A. \& Eichler, D. 1993, ApJ 418, 386

\noindent Levinson, A. \& Eichler, D. 2006, ApJ, 649, L5

\noindent Norris, J. P. and Bonnell, J.T. 2006, ApJ 643, 266.

\noindent Ofek, E. O.; Zoglauer, A.; Boggs, S. E.; Barriere, N. M.; Reynolds, S. P.; Fryer, C. L.; Harrison, F. A.; Cenko, S. B.; Kulkarni, S. R.; Gal-Yam, A.; Arcavi, I.; Bellm, E.; Bloom, J. S.; Christensen, F.; Craig, W. W.; Even, W.; Filippenko, A. V.; Grefenstette, B.; Hailey, C. J.; Laher, R.; Madsen, K.; Nakar, E.; Nugent, P. E.; Stern, D.; Sullivan, M.; Surace, J.; Zhang, W. W., 2013arXiv1307.2247O

 \noindent Rybicki, G.B., Lightman, A. P. 1979, "Radiative Processes in Astrophysics", 1979, (John Wiley and Sons)

\noindent  Ryde, F. 2004, ApJ, 611, L41

 \noindent Sakamoto, et al. 2008, ApJ, 679, 570

 \noindent Thompson, C. 2006, ApJ, 651, 333

\noindent  Vurm, I.,  Lyubarsky, Y., \& Piran, T.  2013, ApJ 764, 143V


   \end{document}